\documentclass[twocolumn,aps,prl]{revtex4}
\usepackage{latexsym}
\usepackage{epsfig}

\def\beq{\begin{equation}}
\def\eeq{\end{equation}}
\newcommand{\bea}{\begin{eqnarray}}
\newcommand{\eea}{\end{eqnarray}}
\def\bi{\begin{itemize}}
\def\ei{\end{itemize}}
\def\Tdot#1{{{#1}^{\hbox{.}}}}

\begin{document}



\title{Evolution of non-linear cosmological perturbations}

\author{David Langlois$^{1,2}$, Filippo Vernizzi$^3$}
\affiliation{$^1$APC (Astroparticules et Cosmologie),
{UMR 7164 (CNRS, Universit\'e Paris 7, CEA, \\Observatoire de
Paris)},
 11 Place Marcelin Berthelot, F-75005 Paris, France}
\affiliation{$^2$Institut d'Astrophysique de Paris,
98bis Boulevard Arago, 75014 Paris, France,}
\affiliation{$^3$Helsinki
Institute of Physics, P.O. Box 64, FIN-00014, University of
Helsinki, Finland}


\vspace{1cm}

\begin{abstract}
We define fully non-perturbative generalizations of the
uniform density and comoving curvature perturbations, which are
known, in the linear theory,  to be conserved on sufficiently
large scales for adiabatic perturbations. Our non-linear
generalizations are defined geometrically, independently of any
coordinate system. We give the equations governing their evolution
on all scales. Also, in order to make contact with previous
works on first and second order perturbations,  we introduce a coordinate
system and show that previous results can be recovered, on large scales, in a
remarkably simple way, after restricting our definitions to first
and second orders in a perturbative expansion.
\end{abstract}

\maketitle

\date{\today}


The relativistic theory of cosmological perturbations is a
cornerstone in  our understanding of the early universe as it is
indispensable in relating the scenarios of the early universe,
such as inflation, to  cosmological data such as the Cosmic
Microwave Background (CMB) anisotropies. This is the reason why
the theory of {\it linear} cosmological perturbations has been
developed to a high degree of sophistication during
 the last twenty-five
years \cite{Bardeen:1980kt,Kodama:1985bj,Mukhanov:1990me}.

With the increase of precision of the  CMB data, the study
of relativistic cosmological perturbations {\it beyond linear
order} is becoming a topical subject, especially to study
primordial non-Gaussianity. This is why much effort has been
devoted recently to the investigation of second-order
perturbations, mainly resorting to an expansion, up to second
order, of Einstein's equations \cite{Acquaviva:2002ud,Noh:2004bc}.
This formalism is rather difficult, due to the complexity of
Einstein's equations,
 and the identification of gauge invariant quantities  is not
 straightforward \cite{Malik:2003mv,Vernizzi:2004nc}.

Other  recent approaches of non-linear perturbations
\cite{rigopoulos,Kolb:2004jg,Lyth:2004gb} are based on the
long wavelength approximation, which at lowest order is related to
the so called separate universe picture that represents our
universe, on scales larger than the Hubble radius, as juxtaposed
Friedmann-Lema\^{\i}tre-Robertson-Walker (FLRW) universes with
slightly different scale factors.

In the present work, we propose a different approach based on a
purely geometrical description of the perturbations. Our approach
 does not rely on any perturbative expansion and is thus not
limited to second-order perturbations. It does not require any
approximation and covers both large and small scales. Furthermore,
as we shall see, it can be related rather easily to the
formulations mentioned above. Our approach is directly inspired by
the so called covariant formalism for cosmological perturbations
introduced by Ellis and Bruni \cite{Ellis:1989jt}. At linear
order, the covariant formalism is {\it computationally} equivalent
to the much more used coordinate approach. However, as we show
here,  it is tremendously more powerful when one goes beyond
linear order.

In particular, we define  geometrically the generalizations of
quantities widely used in the linear theory because they are
conserved on large scales. We then give the full non-linear
equations that govern these quantities. To emphasize the
efficiency of our approach with respect to the more traditional
coordinate-based approach, we show how second order
gauge invariant quantities, recently derived in the literature,
can be recovered, on large scales, from our geometric quantities via a
straightforward derivation.

{\em Covariant approach} -- We consider a spacetime, with metric
$g_{ab}$, filled with a single perfect fluid
 characterized by a  four-velocity $u^a = d x^a/d
\tau$ ($u_a u^a =-1$), where $\tau$ is the proper time along the
flow lines, proper energy density $\rho$ and pressure $P$. The
corresponding energy-momentum tensor is $T^a_{\
b}=\left(\rho+P\right) u^a u_b+Pg^a_{\ b}$. We also introduce the
volume expansion $\Theta \equiv \nabla_a u^a$, which reduces to
the Hubble parameter $H$, $\Theta=3 H$, in a
FLRW spacetime. The
integration of $\Theta$ along the fluid world lines, with respect
to the proper time $\tau$,
\beq
\alpha \equiv {1\over 3}\int d\tau \, \Theta, \quad \quad (\Theta
= 3 \dot\alpha \equiv 3 u^a\nabla_a\alpha ) \label{alpha_def},
\eeq
can be used to define,  for each observer comoving with the fluid,
a local scale factor $e^\alpha$. Note that $\alpha$ is defined up
to an integration constant for {\it each} fluid world line, which
we can set to zero on some reference hypersurface.

As shown by Ellis and Bruni \cite{Ellis:1989jt}, it is possible
to define, in a geometrical way, quantities that can be interpreted
as perturbations with respect to the FLRW configuration, by introducing
spatially projected gradients. The spatial projection tensor,
 orthogonal to the fluid
velocity $u^a$, is defined by $h^a_{\ b}=g^a_{\ b}+u^a u_b$.
Defining the spatial gradient $D_a \equiv h_a^{\ b} \nabla_b$, we
consider the quantities
\bea
X_a\equiv D_a\rho, \quad Y_a\equiv D_a P, \quad Z_a\equiv D_a
\Theta, \quad W_a\equiv D_a \alpha, \nonumber
\eea
which automatically vanish in a strictly FLRW spacetime: in this
sense we call them {\it perturbations}. However, they are fully
non-perturbative quantities. Note that the last quantity, $W_a$,
which as far as we know has not been introduced previously in the
covariant formalism, plays a crucial r\^ole in our approach, where
we replace the familiar spatial  curvature perturbation by the
perturbation of the integrated expansion $\alpha$, i.e., of the
local number of e-folds, in accordance with the separate universe
picture \cite{Sasaki:1995aw,Wands:2000dp}.

{\em Evolution of perturbations} -- \label{sec:evolution} Starting
from the above  definitions, we now derive the fundamental
evolution equation for these non-linear perturbations. 
As in
\cite{Wands:2000dp} in the context of the linear theory,  
our starting point here will be the energy-momentum conservation,
\beq
\label{conserv} \nabla_a T^a_{\ b}=0.
\eeq
The projection along $u^a$ of 
Equation (\ref{conserv}) yields
$$ \dot\rho + \Theta (\rho + P)=0, $$ where a dot represents a
derivative with respect to $\tau$, i.e., stands for $u^a\nabla_a$.
If one takes the projected gradient of this expression one gets, 
after some manipulations,
the relation
\beq
\dot X_a+\left(\rho+P\right)Z_a+\Theta \left(X_a+Y_a\right) +(h^{\
c}_a\nabla_c u^b-\dot h^{\ b}_{a})\nabla_b\rho =0.
\label{first_rel}
\eeq
We then rewrite   $Z_a$ as
\beq
\label{second_rel} Z_a=3\dot W_a+ 3 (h^{\ c}_a\nabla_c u^b- \dot
h^{\ b}_a) \nabla_b\alpha .
\eeq
This suggests introducing the covector
\beq
\label{zeta}
\zeta_a\equiv W_a +{X_a\over {3(\rho+P)}}=D_a\alpha-\frac{\dot\alpha}{\dot\rho}D_a\rho,
\eeq
which represents our {\it non-perturbative generalization
of the curvature perturbation on uniform density hypersurfaces}
or, more exactly, of its ``spatial'' gradient. Interestingly, the same linear combination has
been introduced in the context of the long wavelength approximation,  within a
specific coordinate system, in \cite{rigopoulos}. Here, $\zeta_a$ is
defined geometrically and in the most general context.

The spatial projection of Eq.~(\ref{first_rel}), via $h^{\ b}_a$,  then
yields
\beq
e^{-\alpha}  h_a^{\ b} \Tdot{\left(e^\alpha \zeta_b\right)}=
-{\Theta\over{3(\rho+P)}} \Gamma_{a}-
\zeta^b\left(\sigma_{ba}+\omega_{ba}\right), \label{conserv2}
\eeq
where we have used  the familiar decomposition (see e.g.
\cite{ellis})
$\nabla_b u_a=\sigma_{ab}+\omega_{ab}+{1\over
3}\Theta h_{ab}-\dot{u}_a u_b$ with the
 (symmetric) shear tensor $\sigma_{ab}$, and the
(antisymmetric) vorticity  tensor $\omega_{ab}$;
on the right hand side,
 the quantity
 \beq
 \label{Gamma}
 \Gamma_a\equiv
Y_a -c_s^2 X_a\equiv D_aP- {\dot P\over \dot\rho}D_a\rho
\eeq
represents the non-perturbative generalization of the nonadiabatic
pressure and vanishes for purely adiabatic perturbations, i.e.,
when the pressure $P$ is solely a function of the density $\rho$.
Note also that, for both $\zeta_a$ and $\Gamma_a$, one can replace
the projected derivatives by ordinary partial derivatives
respectively in Eqs.~(\ref{zeta}) and (\ref{Gamma}).

Equation (\ref{conserv2})  is one of the main results of this
work. It gives the fully non-linear evolution equation for the
variable $\zeta_a$ and corresponds to the non-perturbative generalization
of the familiar linear conservation law for adiabatic
perturbations ($\Gamma_a=0$) on large scales, similar to the
results recently obtained in the long wavelength limit \cite{rigopoulos}.

{\em Introducing coordinates} -- \label{coord}
To relate
our approach to the more familiar coordinate-based perturbative
approach, we now introduce a coordinate system, in which the
perturbed metric can be written as $ds^2 =a^2[-(1+2A)d\eta^2 + 2 B_i
dx^i d\eta + \left(\gamma_{ij}+ \delta\gamma_{ij} \right)dx^i
dx^j]$ (see \cite{Bruni:1992dg}).

\def\ab{\bar{\alpha}}
\def\rhob{\bar{\rho}}
\def\chif{{\delta \chi}_{1}}
\def\chis{{\delta \chi}_{2}}
\def\af{{\delta \alpha}_{1}}
\def\as{{\delta \alpha}_{2}}
\def\rhof{{\delta \rho}_{1}}
\def\rhos{{\delta \rho}_{2}}
\def\zetaf{\zeta_1}
\def\zetas{\zeta_2}
\def\Gammaf{\Gamma_1}
\def\Gammas{\Gamma_2}
\def\Rf{{\cal R}_1}
\def\Rs{{\cal R}_2}
\def\psif{\psi_1}
\def\psis{\psi_2}
\def\vf{v_1}
\def\vs{v_2}
\def\Bf{B_1}
\def\Bs{B_2}

In the following, we decompose any quantity $\chi$  in the form
$\chi (\eta,x^i)={\bar \chi}(\eta)+ \chif(\eta,x^i)+\frac{1}{2}
\chis(\eta,x^i)$, where $\chif$ and $\chis$ represent,
respectively, the first and second order perturbations. Let us
write down explicitly the components of our covector $\zeta_a$ in
this coordinate system.  The
 zeroth order vanishes and, at {\it first order}, the spatial
components are simply
\beq
\zeta_{i}^{(1)}=\partial_i \zetaf, \qquad \zetaf\equiv
\af-{{\ab}'\over {\rhob}'} \rhof, \label{zeta1}
\eeq
where a prime denotes the partial derivative with respect to the
time coordinate $\eta$. To compute the component $\zeta_0$, it is
useful to note that for any function $\chi$, one can write
$D_0\chi=u_0 u^i\partial_i\chi-u^iu_i\partial_0 \chi$, where we
have used the normalization of $u^a$. Since $u^i$ is first order,
this implies that $\zeta_0^{(1)}=0$ and
$\zeta_0^{(2)}=u^i\zeta_i^{(1)}$.

Expanding $\zeta_i=\partial_i
\alpha-(\dot\alpha/\dot\rho)\partial_i\rho$
 {\it at second order}, one finds, after some simple manipulations,
\beq
\label{zeta2} \zeta_i^{(2)}=\partial_i\zetas + {2\over
\rhob'}\rhof\partial_i \zetaf{}',
\eeq
\beq
\zetas \equiv \as - {{\ab}'\over {\rhob}'}\rhos
 - {2\over \rhob'}\af'\rhof+2{{\ab}'\over {{\rhob}'}{}^2}\rhof\rhof'+{1\over \rhob'}
 \left({{\ab}'\over {\rhob}'}\right)'
 \rhof^2.
  \label{zeta_scal}
\eeq
Note that the conservation of
 $\zeta_i^{(1)}$ and  $\zeta_i^{(2)}$  is equivalent
to the conservation of $\zetaf$ and $\zetas$.

We now relate $\alpha$ to the metric perturbations, considering
for simplicity only large scales. Neglecting gradients as well as first order
vector and tensor perturbations,  and writing
$\psi\equiv-\gamma^{ij}\delta\gamma_{ij}/3$, one
 can  use the equality $3 \dot \alpha=
\nabla_a u^a$ to derive, {\em  up to second order}, the relation
\bea
\alpha \simeq \ln a  -\psi-\psi^2.
\eea
Thus, from Eq.~(\ref{zeta_scal}), we can relate $\zetas$ to the
quantity defined  by Malik and Wands in \cite{Malik:2003mv},
\bea
\zetas \simeq \zeta_{2\rm (MW)} - 2\zeta^2_{1\rm (MW)}. \nonumber
\eea

The  expansion of the nonadiabatic term $\Gamma_a=\partial_a
P-(\dot P/\dot\rho)\partial_a\rho$, which can be read from the
expansion of $\zeta_a$ by substituting $P$ to $\alpha$, enables us
to write explicitly the conservation equation (\ref{conserv2}) at
first and second orders. If one ignores the second term on the
right hand side of Eq.~(\ref{conserv2}), one finds,
at first and second order, respectively,
\bea
\zetaf' &\simeq& - \frac{{\cal H}}{\bar \rho + \bar P} \Gammaf,
\nonumber \\
\zetas'  &\simeq& - \frac{{\cal H}}{\bar \rho + \bar P} \Gammas -
\frac{2}{\bar \rho + \bar P} \Gammaf  \zetaf{}', \nonumber
\eea
where the $\Gamma_{1,2}$ are defined as the $\zeta_{1,2}$, and
${\cal H} \equiv a'/a$. One thus recovers the results of
\cite{Malik:2003mv} very easily. Moreover, Eq.~(\ref{conserv2})
generalizes them to any order in the perturbation theory and to
all scales.

{\em Comoving ``curvature'' perturbation} -- In addition to the
uniform energy density curvature perturbation, generalized by our
$\zeta_a$, another useful quantity in the linear theory is the
curvature perturbation on comoving hypersurfaces, usually denoted
by ${\cal R}$. In our formulation, since the perturbed integrated
expansion $\alpha$ replaces the usual spatial curvature and since
 our spatial gradients are defined with respect to the
comoving observers, the generalization of ${\cal R}$ is simply
\beq
{\cal R}_a \equiv -D_a \alpha = -W_a.
\label{R}
\eeq
In the coordinate system introduced earlier, the spatial
components are
\bea
{\cal R}_i=-\partial_i \alpha- \dot\alpha a(v_i+B_i), \nonumber
\eea
with $v^i\equiv a u^i$.
At first order, $\delta \alpha\simeq -\psi$, and one recognizes
the familiar comoving curvature perturbation of the linear theory.

It is relatively easy to derive the evolution equation for ${\cal
R}_a$. Equation (\ref{second_rel}) can be seen as an equation for
$\dot{\cal R}_a= -\dot{W}_a$. Projecting orthogonally to  $u^a$,
we find
\beq
h_a^{\ b} \dot{\cal R}_b+ \frac{1}{3} \Theta {\cal R}_a= - {\cal R}^b
(\sigma_{ba} +\omega_{ba}) - h_a^{\ b} \dot u_b \dot\alpha
-\frac{1}{3} Z_a \label{step1}.
\eeq
The term involving  $Z_a$ on the right hand side can be rewritten
by using the shear constraint equation [see Eq.~(4.17) of
\cite{ellis}],
\beq
\frac{2}{3} Z_a - h_a^{\ b} \nabla_c (\sigma^{c}_{\ b} + \omega^{
c}_{\ b}) + (\sigma^{\ b}_{a} + \omega^{\ b}_{a}) \dot u_b= q_a = 0,
\label{Za}
\eeq
 where $q_a$ is the energy
flux in the energy-momentum tensor, which, in our case, vanishes.
Moreover, one can replace $\dot u_b$ using the Euler equation, $\dot u_a =
-Y_a/(\rho+P)$. On
rewriting $Y_a$ as $Y_a = \Gamma_a+ c_s^2 X_a$, and combining all
this into Eq.~(\ref{step1}), we finally obtain
\bea
e^{-\alpha} h_a^{\ b} \Tdot{\left(e^\alpha {\cal R}_b \right)} =
\left[\frac{\Theta \delta^{\ b}_a}{3(\rho+ P)} - \frac{\sigma^{\
b}_{a}+\omega^{\ b}_{a}}{2(\rho+P)} \right] \left( \Gamma_b +
c^2_s X_b \right) \nonumber
\\
-{\cal R}^b(\sigma_{ba} +\omega_{ba}) -\frac{1}{2} h_a^{\
b} \nabla_c(\sigma^{c}_{\ b} + \omega^{
c}_{\ b}).   \label{Revol}
\eea
This is the fully non-linear evolution equation for ${\cal R}_a$.
As one can see, it is slightly more complicated than the evolution
equation for $\zeta_a$. However,  when
$\sigma_{ab}$ and $\omega_{ab}$ are negligible, one can
recover the familiar conservation law for adiabatic perturbations
by noting that $X_a$, which generalizes the comoving energy
density perturbation of the linear theory \cite{Kodama:1985bj}, can be
shown to be
negligible on large scales.
Since $\zeta_a +{\cal R}_a = -({\dot\alpha}/{\dot\rho}) X_a$, one then
finds $\zeta_a \simeq - {\cal R}_a $ non-perturbatively.

Note that ${\cal R}_a$ is not the only covariant quantity that can
be connected, within a coordinate system,  to the familiar
comoving curvature perturbation of the linear theory. This is also
the case for the projected gradient of the spatial scalar
curvature (denoted by $C_a$ in, e.g., \cite{Bruni:1992dg}),
although there does not seem to be a simple relation at the
non-perturbative level between ${\cal R}_a$ and $C_a$.

{\em Scalar field} -- It is also straightforward to extend  our
non-perturbative approach to a scalar field. Let us  consider a
scalar field $\phi$ with a potential $V(\phi)$. Its
energy-momentum tensor, $T_{ab} =
\partial_a \phi \partial_b \phi - \frac{1}{2}g_{ab} \left(
\partial_c \phi \partial^c \phi +2 V \right)$,
can be rewritten in the perfect fluid form  by defining the
four-velocity $u^a$ as  the unit vector orthogonal to
hypersurfaces of constant $\phi$ \cite{Bruni:1991kb},
\beq
\label{u} u_a\equiv -{\partial_a \phi}/{(-
\partial_c \phi \partial^c \phi)^{1/2}}=
-{\partial_a \phi}/{\dot\phi},
\eeq
where the second equality (with $\dot\phi=u^a\partial_a\phi$) is a
consequence of the definition of $u^a$, and the energy density and
pressure as
$\rho =  \dot\phi^2/2 + V$ and $P =\dot\phi^2/2 - V$.

For a scalar field, the comoving curvature perturbation is
particularly convenient because the $\phi=const$ hypersurfaces are
orthogonal to $u^a$ defined in (\ref{u}). Since $D_a\phi=0$, one
sees that  ${\cal R}_a=-D_a\alpha$ can be reexpressed as
\bea
{\cal R}_a= -\partial_a \alpha + \frac{\dot\alpha}{\dot\phi}
\partial_a \phi \nonumber,
\eea
and one recognizes the comoving curvature perturbation of the linear theory.
The generalization of the Mukhanov variable \cite{Mukhanov:1990me}
for the quantization of the scalar field-gravity system
in the non-linear theory is
thus $v_a\equiv e^\alpha (\dot \phi/\dot \alpha){\cal R}_a$.

In a coordinate system, ${\cal R}_a$ can be easily expanded to
first and second orders to make contact with previous results. As
for $\zeta_a$, one finds, at first and second orders, expressions
of the form
\bea
{\cal R}^{(1)}_i = \partial_i \Rf, \quad {\cal R}^{(2)}_i =
\partial_i \Rs+ {2\over \bar{\phi}'} \delta
\phi_{1} \partial_i \Rf', \nonumber
\eea
where the quantities $\Rf$ and $\Rs$ can be deduced from $\zetaf$
and $\zetas$, given in (\ref{zeta1}) and (\ref{zeta2}), by
changing the overall sign and replacing $\rho$ by $\phi$.
When first order vector and tensor modes can be neglected, up to
small scale terms and an overall sign, ${\cal R}={\cal
R}_1+\frac{1}{2}{\cal R}_2$ coincides with the variables used in
\cite{Maldacena:2002vr} (see also \cite{Salopek}) either in  the
uniform field gauge, which can be obtained by setting $\delta
\phi=0$, or the uniform curvature gauge, obtained by setting
$\delta \alpha=0$. On writing $\alpha$ in terms of the
metric components, and taking the large scale limit, one can also
show that $\cal R$ can be simply related to the conserved
quantities defined in \cite{Noh:2004bc,Vernizzi:2004nc}.

For a scalar field ($\omega_{ab}=0$),  if $\sigma_{ab}$
can be neglected, one can also generalize the well known first order
property that the nonadiabatic term $\Gamma_a$ is
negligible  in the large scale limit. This  result, which extends
what  is known up to second order \cite{Vernizzi:2004nc}, is
obtained by noting that $D_a V=0$ (since $D_a\phi=0$), which
implies $\Gamma_a= (1-c_s^2) X_a \simeq 0$. In this case, for a {\em
single scalar field}, the right hand side of Eq.~(\ref{Revol}) is
negligible and ${\cal R}_a$ is conserved, on
super-Hubble scales, to all orders in perturbation theory. If more
fields are present, then $\Gamma_a \simeq -2 D_a V \neq 0$ and the
non-linear perturbation ${\cal R}_a$, and thus the
non-Gaussianity, can evolve on large scales during inflation.

{\em Conclusion} -- In this Letter, we have defined,
respectively in (\ref{zeta}) and (\ref{R}), non-perturbative
generalizations of the so called uniform energy density curvature
perturbation and comoving curvature perturbation, using the
integrated expansion $\alpha$, which represents the number of
e-folds of the local expansion measured by an observer comoving
with the fluid.

Although our perturbed quantities are similar to those advocated
in recent works based on the long wavelength approximation
\cite{rigopoulos,Kolb:2004jg,Lyth:2004gb}, their originality is
that they are constructed geometrically, independently of any
coordinate system. Furthermore, we have been able to derive
evolution equations governing these quantities, which are  {\it
fully non-linear} and {\it  exact at all scales}.
 In particular,
Eq.~(\ref{conserv2}), obtained directly from the local
conservation of the energy-momentum tensor and independently of
Einstein's equations, is  valid in any metric theory of gravity,
including scalar-tensor theories or induced four-dimensional
gravity in brane-world scenarios. Thus, our
relations generalize the familiar conservation laws of the linear
theory for adiabatic perturbations.

Another advantage  of our formulation is that it is rather easy to
relate to   the coordinate-based perturbative formulation.
Interestingly, our extremely simple non-linear expressions, like
(\ref{zeta}), automatically contain the seemingly complicated
second-order structure that has been obtained in the
coordinate-based approach,
which we can 
easily
 rederive and extend to
higher orders.

Finally, let us stress that our formulation, since it keeps track
of all small scale terms, may simplify the study of non-linear
perturbations even inside the Hubble radius. This would be very
useful, in particular to compute  the non-Gaussianity from
inflation \cite{Maldacena:2002vr}  or to study the backreaction of
small scale modes on larger scales.

We thus believe that this approach might represent a significant
contribution to the study of primordial cosmological perturbations
beyond the linear order.

{\em Acknowledgments:} We thank Marco Bruni for crucial comments
on a preliminary version of this work and Misao Sasaki for useful
discussions. F.V. thanks the Institut d'Astrophysique de Paris for
kind hospitality.


\end{document}